\documentclass[useAMS,usenatbib]{mn2e}

\usepackage{graphicx}

\title[2-D models of layered protoplanetary discs]{Two-dimensional models of layered
protoplanetary discs -- II.\\The effect of a residual viscosity in the dead zone.}
\author[R.~W\"unsch, A.~Gawryszczak, H.~Klahr and M.~R\'o\.zyczka]
{R.~W\"unsch$^{1,3}$,\thanks{E-mail: richard.wunsch@matfyz.cz}
A.~Gawryszczak$^{1}$, H.~Klahr$^{2}$ and M.~R\'o\.zyczka$^{1}$\\
$^{1}$Nicolaus Copernicus Astronomical Center, Bartycka 18, 00-716 Warsaw, Poland\\
$^{2}$Max-Planck-Institut f\"ur Astronomie, K\"onigstuhl 17, D-69117 Heidelberg, Germany\\
$^{3}$Astronomical Institute, Academy of Sciences of the Czech
Republic, Bo\v{c}n\'\i\ II 1401, 141 31 Prague, Czech Republic}
\begin{document}

\date{Received: }

\pagerange{\pageref{firstpage}--\pageref{lastpage}} \pubyear{2005}

\maketitle

\label{firstpage}

\begin{abstract} % {{{
We study axisymmetric models of layered protoplanetary discs
taking radiative transfer effects into account, and allowing for a
residual viscosity in the dead zone. We also explore the effect of
different viscosity prescriptions. In addition to the ring
instability reported in the first paper of the series we find an
oscillatory instability of the dead zone, accompanied by
variations of the accretion rate onto the central star. We provide
a simplified analytical description explaining the mechanism of
the oscillations. Finally, we find that the residual viscosity
enables stationary accretion in large regions of layered discs.
Based on results obtained with the help of a simple 1-D hydrocode
we identify these regions, and discuss conditions in which layered
discs can give rise to FU~Orionis phenomena.
\end{abstract}
% }}}


\section{Introduction} % {{{

The transport of angular momentum in protoplanetary discs is most
probably driven by the magnetorotational instability, described by
\citet{bh91a,bh91b}, and hereafter referred to as MRI.
Since the protoplanetary discs are ionized rather weakly, in some
regions of the disc the ionization degree $\xi$ can decrease below
a critical level, $\xi_c$, at which the gas decouples from the
magnetic field and the MRI decays. Such region is referred to as a
{\em dead zone}, contrary to the so called {\em active zone} where
the MRI operates.

The most important processes responsible for the ionization of gas
in protoplanetary discs are 1) particle collisions due to thermal
motions, and 2) irradiation by cosmic rays and high-energy photons
originating from the central star. The collisional ionization
dominates in the innermost part of the disc, where the temperature
is high enough to maintain $\xi>\xi_c$ at all distances from the
disc plane. Further away from the star $\xi_c$ is exceeded only in
two surface layers (on both sides of the disc), each ionized by
cosmic rays and stellar X-ray quanta, and having approximately
constant column density $\Sigma_\mathrm{a}$. Both processes result in the
following distribution of disc activity: at $r < r_1$ (where the
thermal ionization degree exceeds $ \xi_c$) the whole disc is
active; at $r_1<r<r_2$ (where $\Sigma>\Sigma_\mathrm{a}$) a dead zone is
sandwiched between two active layers; and at $r > r_2$ (where
$\Sigma< \Sigma_\mathrm{a}$) the whole disc becomes active again. This
model was proposed by \citet{gammie96} to explain the FU~Ori
outbursts.

FU~Ori stars are young stellar objects whose optical brightness
occasionally increases by $3-4$~mag on a time-scale of $1-10$~yr,
and remains at the increased level for $10-100$~yr. It is
generally believed that these events are related to the variations
of the accretion rate in circumstellar discs. To explain the
observed increase in luminosities the accretion rate has to change
from $\sim 10^{-7}$ to $\sim 10^{-4}$~M$_\odot$yr$^{-1}$, implying
that as much as $10^{-2}$M$_\odot$ must be accreted onto the
central star during the outburst \citep{hk96}.

The FU~Ori phenomenon may be caused by a thermal instability of the
protoplanetary disc (see e.g. \citet{lc04}, \citet{bl94} and
references therein), analogous to the one operating in accretion discs
around primary components of cataclysmic binaries \citep{mm-h81}.
However, the thermal instability model has difficulties with
explaining long duration of the outbursts, since the high accretion
rate during the outburst quickly removes most of the material from the
inner part of the disc where the instability occurs \citep{hk96}.
A transition of the very inner disc to the outburst state due to the
thermal instability was studied by \citet{kl99} with the help of the
similar two-dimensional code as used in this work.

\citet{gammie96} suggested that this problem may be solved by the
layered model, in which the dead zone serves as a mass reservoir
for the outburst. In the layered disc the accretion rate can be an
increasing function of $r$. When this is the case, an annulus
centred at $r_0$ receives more mass per unit time from the disc
exterior to it $(r>r_0$ than it loses to the disc interior to it
$(r<r_0)$, causing the matter to accumulate in the dead zone. If
this process lasts long enough, the total surface density of the
disc may become so large that once the accretion was somehow
triggered in the dead zone it would release enough heat to restore
the coupling between the matter and the magnetic field. Such an
"ignition" of the accumulated matter would then start a
self-sustaining outflow of mas from the dead zone into the inner
disc, substantially increasing the accretion rate.

\citet{alp01} assume that the triggering factor is the gravitational
instability. Their model shows repeating periods of vigorous accretion
separated by quiescent intervals lasting $\sim 10^5$~yr, in
qualitative agreement with the observed properties of FU~Ori objects.
However, our radiation-hydrodynamic simulations of layered discs
(\citealt*{wkr05}, hereafter Paper~I) indicate, that it may be rather
difficult for the dead zone to store $\sim10^{-2}$M$_\odot$ of the
disc matter necessary to feed an FU~Ori outburst. The reason is the
ring instability described in Paper~I, which, long before the required
mass is reached, causes the dead zone to split into well-defined rings
that become unstable to the instability discovered by \citet{pp85}.

In the present paper we extend the simulations reported in Paper~I
onto a broader range of disc parameters, with the principal aim to
find in which conditions the dead zone could become massive enough to
account for FU~Ori outbursts. For reasons explained in \S2, along with
models whose dead zones are entirely free of viscosity we explore
models with some residual viscosity acting in the region which
formally should be declared dead. We also explore the effect of
different viscosity prescriptions.

The paper is organized as follows. In \S2 we briefly describe the
numerical methods and input physics. The results of simulations
are reported in \S3. In \S4 we give a simple analytical
description of processes at the inner boundary of the dead zone,
and derive conditions for ignition. In \S5 we summarize our
results and discuss the limits of the mass accumulation process.

% }}}

\section{Numerical methods and input physics} % {{{

As in Paper~I, the simulations are performed in spherical
coordinates with the help of the radiation-hydrodynamics code
TRAMP \citep*{khk99}. The models are axially symmetric, but
we do not impose an additional mirror symmetry with respect to the
mid-plane of the disc, i.e. we allow for the flow through the
mid-plane. Further technical details, like boundary conditions or
initialization procedure, can be found in Paper~I.

The viscosity is treated according to the $\alpha$-prescription of
\citet{ss73}. In Paper~I, following the argument that the turbulent viscosity
coefficient should be proportional to the product of sound speed and the
largest scale available for the turbulent eddies, we used the formula

\begin{equation}
\nu = \alpha c_\mathrm{s,i} H_\mathrm{a} \ ,
\end{equation}
where $H_\mathrm{a}$ is the half-thickness of the active layer, $c_\mathrm{s,i}$ is the
sound-speed at the bottom of the active layer and $\alpha$ is a
dimensionless parameter discussed below. In the present work this
formula is employed in models $1-3$. The remaining models are
obtained based on an alternative formula which allows for
variations of $\nu$ with the distance from the mid-plane
\begin{equation}
\nu = \alpha \frac{c_\mathrm{s}^2}{\Omega} \ ,
\end{equation}
where $c_\mathrm{s}$ and $\Omega$ are {\it local} values of sound speed and
angular velocity. Obviously, in a vertically averaged
approximation assuming the zero-thickness dead zone both prescriptions
are equivalent; however, in two-dimensional models with the dead zone,
we found that the details of the evolution are sensitive to the
particular form of the viscosity formula (see \S3).

We assume that at $T > T_\mathrm{lim} = 1000$~K the ionization
degree is high enough for the magnetorotational instability to
operate and provide the source of viscosity, whereas at $T <
T_\mathrm{lim}$ the MRI is quenched. This assumption is based on
the fact that at $T \sim 1000$~K $\xi$ increases by five orders of
magnitude due to ionization of potassium \citep{umebayashi83}.
Further, based on the cosmic ray stopping depth calculated by
\citet{un81}, we assume that on both sides of the disc a surface
layer of column density $\Sigma_\mathrm{a} = 100$~g\,cm$^{-2}$ is made
active due to ionization by energetic particles of cosmic origin.

In principle, ionization due to X-rays from the central star should
also be taken into account. However, the corresponding active surface
layer is probably not thicker than a few tens of g\,cm$^{-2}$
\citep*{gni97}. The X-rays become important when the disc is screened
from cosmic rays by magnetized stellar winds \citep{gni97,ftb02}. In
such cases the ionization structure of the disc strongly depends on
various model parameters like properties of the X-ray source, critical
magnetic Reynolds number or abundance of heavy metals \citep{ftb02},
and it is not possible to define one critical $\Sigma_\mathrm{a}$ for all
possible circumstances. Unfortunately, such effects are too
complicated to take them into account in multidimensional simulations,
and an active surface layer of constant column density must serve as
the necessary approximation. In effect, to define the boundary of the
dead zone we employ the simple model of disc ionization structure
introduced by \citet{gammie96}.

In Paper~I $\alpha$ was set to 0 within the dead zone. However, it has
been recently argued that the dead zone is not entirely dead in the
sense that a residual viscosity can be maintained there by
a purely hydrodynamic turbulence, excited by the non-axisymmetric
waves propagating from the MRI-turbulent active layers \citep{fs03}.
Following those arguments, in some models we allow for
$\alpha_\mathrm{DZ}\ne
0$. However, in all models the dead zone is much less viscous than the
active zone, i.e. $\alpha_\mathrm{DZ} \ll \alpha_\mathrm{a}$.

In the previous work we neglected the transport of heat by micro-turbulence,
i.e. the heat was transported across the disc by radiation only. The
same approximation is adopted here. While it would be poorly justified
in 1-D models (see e.g. \citealt{cannizzo01}), it is entirely
reasonable in 2-D discs, where the most important (i.e. the largest)
turbulent scales are automatically taken into account, as they are
resolved on the grid. Based on the similarity of the turbulence
maintained by the magnetorotational instability to Kolmogoroff
turbulence \citep{hawley95,stone96}, we estimated that the subgrid
micro-turbulent flux is not significant compared to the radiative one.

% }}}

\section{Results} % {{{

Varying $\alpha_\mathrm{a}$, $\alpha_\mathrm{DZ}$ and the viscosity prescription, we
obtained a set of disc models listed in Table \ref{modeltab}. Models
1-5, 7 and 8 were followed for several thousand orbits of the outer
edge of the disc on a grid of $128\times32$ points in $r$ and
$\theta$, respectively. Model 6, which is compared to the analytical
description of the disc obtained in \S\ref{sec:oscil} is computed on a
grid with double resolution ($256\times64$). The size of the
computational domain was set so that it included a small part of the
inner $\alpha$-disc and a part of the dead zone where the ignitions
took place. In radial direction, it was $r=(0.05, 0.35)$~AU for
models 1-6, $r=(0.04,0.3)$~AU for model 7 and $r=(0.1,0.7)$~AU for
model 8. The vertical extent of the computational domain was $\theta =
(-5^o, 5^o)$ for all models.

\begin{table}
\begin{tabular}{ccccc}
Model & $\nu$-type        & $\alpha_\mathrm{a}$         & $\alpha_\mathrm{DZ}$ & grid       \\
\hline
1 & $\alpha c_\mathrm{s} H_\mathrm{a}$ & $10^{-2}$          & $0$               & $128\times32$ \\
2 & $\alpha c_\mathrm{s} H_\mathrm{a}$ & $10^{-2}$          & $10^{-4}$         & $128\times32$ \\
3 & $\alpha c_\mathrm{s} H_\mathrm{a}$ & $10^{-2}$          & $10^{-3}$         & $128\times32$ \\
4 & $\alpha c_\mathrm{s}^2/\Omega$     & $10^{-2}$          & $0$               & $128\times32$ \\
5 & $\alpha c_\mathrm{s}^2/\Omega$     & $10^{-2}$          & $10^{-4}$         & $128\times32$ \\
6 & $\alpha c_\mathrm{s}^2/\Omega$     & $10^{-2}$          & $10^{-3}$         & $256\times64$ \\
7 & $\alpha c_\mathrm{s}^2/\Omega$     & $5\times 10^{-3}$  & $5\times 10^{-4}$ & $128\times32$ \\
8 & $\alpha c_\mathrm{s}^2/\Omega$     & $2\times 10^{-2}$  & $2\times 10^{-3}$ & $128\times32$ \\
\end{tabular}
\caption{Parameters of the disc models.} \label{modeltab}
\end{table}
\subsection{Model 1. ($\nu=\alpha c_\mathrm{s} H_\mathrm{a}$, $\alpha_\mathrm{a}=10^{-2}$, $\alpha_\mathrm{DZ} = 0$)}
This model is the same as the one described in Paper~I, but the
current resolution is two times lower. The main purpose of this
calculation was to check if we recover all essential features and
time-scales observed and measured in Paper~I, which is indeed the
case. The model develops a strong instability, which quickly leads
to the formation of well-separated rings, most of which remain
stable for at least a few hundred years. However, they would probably
decay in 3D due to the hydrodynamic instability discovered by
\citet{pp85} as noted in Paper~I.

Paper~I contains a detailed discussion of the ring
instability mechanism. Here we only mention that the key element
of that mechanism is the coupling between the changes of the dead
zone thickness and the mass accumulation rate. It turns out that
the accretion rate drops at the inner edge of the ring, which
works as a bottle-neck in the accretion flow along the active
surface layer. As a consequence, matter accumulates preferably in
the ring.

Since only a small amount of mass flows from the ring inwards,
the growth of a given ring is substantially slowed down when
additional rings are formed further away from the centre of the
disc. If several rings are simultaneously present in the grid, the
accretion rate (always measured at the inner boundary of the
computational domain) drops to a value as low as $\sim 3\times
10^{-9}$~M$_\odot$yr$^{-1}$. As the ring accretes mass and becomes
denser, its central temperature increases. Once it exceeds
$T_\mathrm{lim}$, the ring ignites and a "mini-outburst" ensues,
after which the ring is rebuilt at the same position. The
accretion rate onto the star increases up to to approximately
$5\times 10^{-9}$~M$_\odot$yr$^{-1}$ during an ignition of a
single innermost ring (at $t=80$ and $220$~yr). Moreover, two
cases of "induced ignition" are observed at $t = 490$ and
$680$~yr. In such a case the heat produced by the ignition and
accretion of a particular ring is transported by the radiation to
the neighbour rings and makes them active. The accretion rate onto
the star increases up to $1.4$ and $1.0\times
10^{-8}$~M$_\odot$yr$^{-1}$.

\subsection{Model 2. ($\nu=\alpha c_\mathrm{s} H_\mathrm{a}$, $\alpha_\mathrm{a}=10^{-2}$, $\alpha_\mathrm{DZ} = 10^{-4}$)}

This model evolves very similarly to Model~1. The non-zero
viscosity in the dead zone causes the temperature inside the rings
to increase faster, and the induced ignition occurs earlier than in
Model~1. The accretion rate in both quiet and ignited states are
almost the same as in Model~1.

\subsection{Model 3. ($\nu=\alpha c_\mathrm{s} H_\mathrm{a}$, $\alpha_\mathrm{a}=10^{-2}$, $\alpha_\mathrm{DZ} = 10^{-3}$)}

In this model we also observe the tendency toward ring formation. However, the
viscosity in the dead zone is now sufficiently high to prevent it from
splitting. The behaviour characteristic of the rings in Models 1 and 2 is now
observed in the whole low-viscosity area: as the surface density of the disc
grows, the temperature at the mid-plane grows as well, and ignition occurs once
it exceeds $T_\mathrm{lim}$. The accretion rate is increased to $\sim 1.3\times
10^{-8}$~M$_\odot$yr$^{-1}$, the accumulated mass is quickly removed from the
ignited region, and the layered structure is re-established. Such
mini-outbursts repeat on a time-scale of $100-200$~yr. They are not strictly
periodic, since the surface density of the layered part of the disc is not
completely smooth, as it is incessantly perturbed by the ring instability. At
$t=700$~yr the whole dead zone contained in the computational domain is ignited
and the accretion rate increases up to $4.7\times 10^{-8}$~M$_\odot$yr$^{-1}$.
Since the mini-outburst is stopped by the outer boundary, and the peak
accretion rate is probably underestimated.

\subsection{Model 4. ($\nu=\alpha c_\mathrm{s}^2/\Omega$, $\alpha_\mathrm{a}=10^{-2}$,
$\alpha_\mathrm{DZ} = 0$)}
Simulation that produced Model~1 is now repeated with the modified viscosity
prescription. Since in the modified prescription there is no direct connection
between the thickness of the active layer and the viscosity coefficient, the
ring instability needs much more time to develop.
The growth rate of the instability is approximately one order of
magnitude lower than in Model 1. By the end of the simulation only the
innermost ring is separated from the dead zone, while the
remaining ones are visible only as density enhancements within the
dead zone. No ignition events are observed. The radial extent of
the innermost ring is approximately twice as big as the one in
Model~1, and its mass is $5\times 10^{-6}$~M$_\odot$
(approximately five times more than in Model~1).

\subsection{Model 5. ($\nu=\alpha c_\mathrm{s}^2/\Omega$, $\alpha_\mathrm{a}=10^{-2}$,
$\alpha_\mathrm{DZ} = 10^{-4}$)}
This model differs from Model~2 in the same way as Model~4 from Model~1, i.e.
only in the formula employed to describe the viscosity. With the modified
formula the residual viscosity in the formally dead area is high enough to
prevent the formation rings before the dead zone ignites at $\sim 700$~yr. As a
result of ignition the accretion rate increases $5\times
10^{-9}$~M$_\odot$yr$^{-1}$ to $3\times 10^{-8}$~M$_\odot$yr$^{-1}$. This
mini-outburst involves the whole dead zone in the computational domain, so the
peak accretion rate is probably underestimated, as in the case of Model~3.
Approximately $5\times 10^{-6}$~M$_\odot$ were accreted onto the star during
this mini-outburst.

\subsection{Model 6. ($\nu=\alpha c_\mathrm{s}^2/\Omega$, $\alpha_\mathrm{a}=10^{-2}$,
$\alpha_\mathrm{DZ} = 10^{-3}$)}
In this case, the high viscosity in the dead zone completely
suppresses the ring instability and the surface density remains
smooth in the layered region. Recurrent ignitions and
replenishments of the dead zone lead to the periodic oscillations
of its inner boundary. The period of these oscillations is around
$30$~yr. Since the mass accumulates in the whole dead zone, a
progressively larger part of it is ignited at each oscillation,
and the amplitude of the oscillations generally increases. This
increase is accompanied by a slow secular increase of the
accretion rate at the inner boundary (from $1\times
10^{-8}$~M$_\odot$yr$^{-1}$ in the beginning to $1.3\times
10^{-8}$~M$_\odot$yr$^{-1}$ at $364$~yr).
After each ignition the accretion rate temporarily increases by
approximately $10\%$.

\begin{figure*}
\noindent
\includegraphics{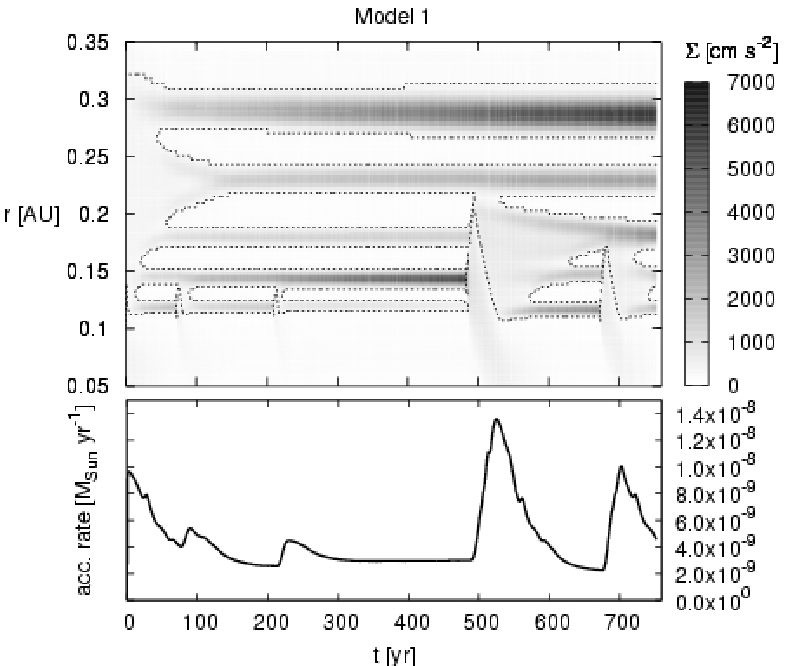}
\includegraphics{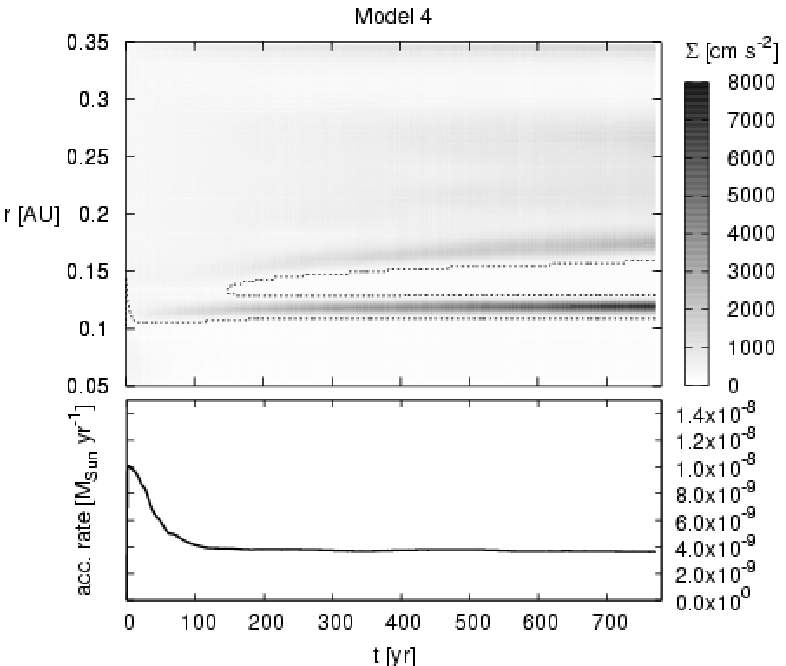}

\includegraphics{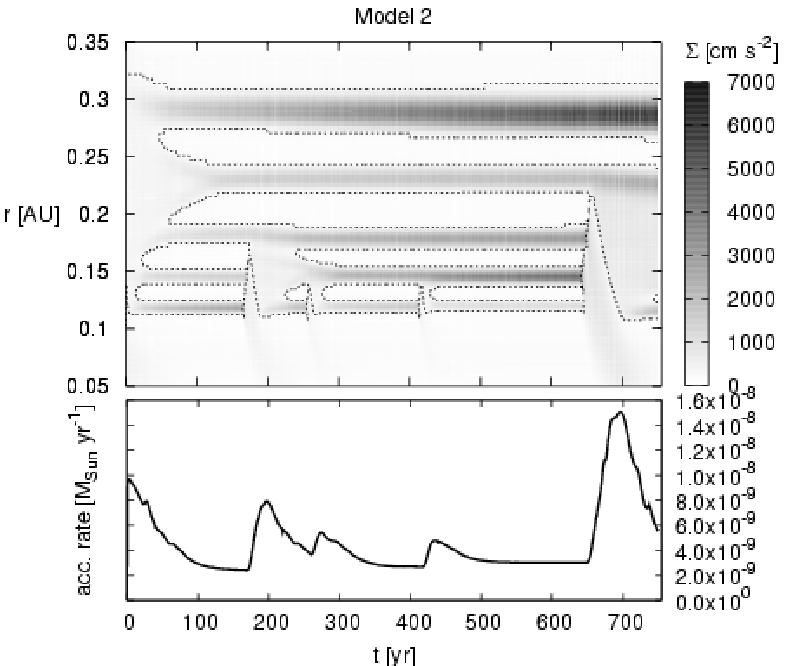}
\includegraphics{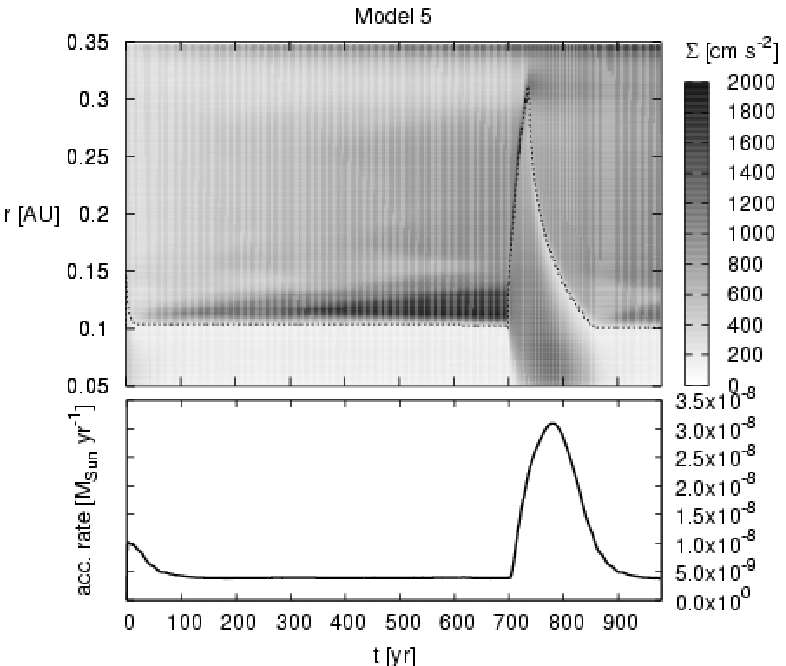}

\includegraphics{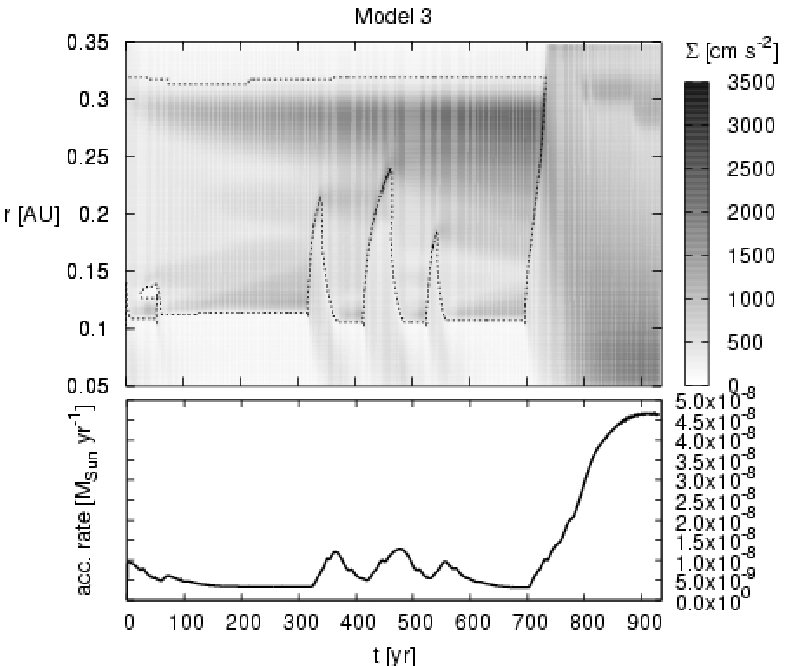}
\includegraphics{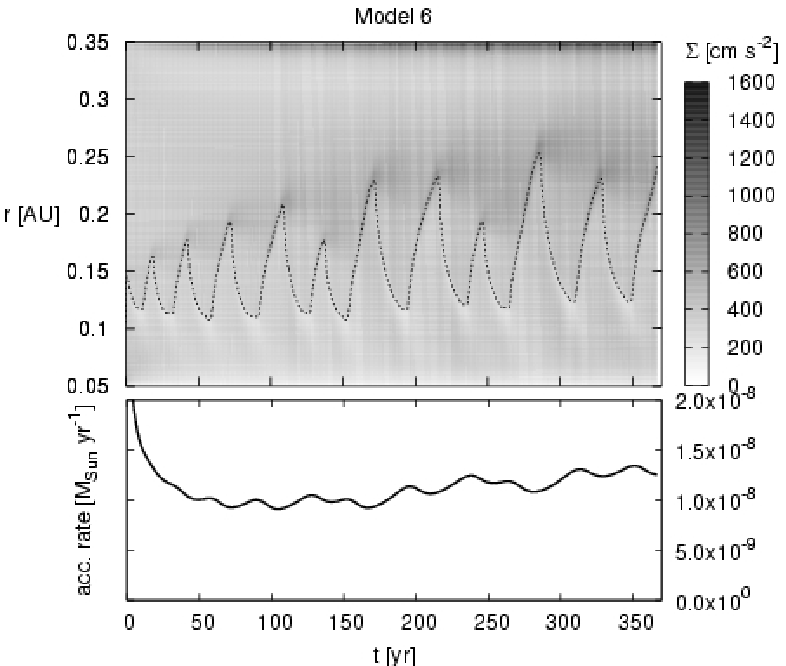}
%\vspace*{22cm}
\caption{Models 1-6: the top figures show the evolution of the disc
surface density (levels of grey), the dashed line denotes the boundary
between the dead zone and the active parts of the disc in the
mid-plane. The bottom figures show the accretion rate at the inner
boundary of the computational domain.}
\label{ar}
\end{figure*}

\subsection{Model 7. ($\nu=\alpha c_\mathrm{s}^2/\Omega$, $\alpha_\mathrm{a}=5\times 10^{-3}$, $\alpha_\mathrm{DZ} = 5\times 10^{-4}$)}

The model evolves similarly to Model 6. The accretion rate slowly
grows from $2\times 10^{-9}$ in the beginning to $5.5\times
10^{-9}$~M$_\odot$yr$^{-1}$ at $t=600$~yr. The relative changes
during mini-outbursts are slightly higher, reaching approximately
$30\%$. At $t=400$~yr a strong convection occurs in the inner
$\alpha$-disc which makes the accretion rate curve noisy.

\subsection{Model 8. ($\nu=\alpha c_\mathrm{s}^2/\Omega$, $\alpha_\mathrm{a}=2\times 10^{-2}$, $\alpha_\mathrm{DZ} = 2\times 10^{-3}$)}

This model exhibits very regular mini-outbursts with period $\sim
150$~yr. The accretion rate at the inner boundary changes from
$2\times 10^{-8}$ in the quiet state to the peak value $6\times
10^{-8}$~M$_\odot$yr$^{-1}$. The amplitude of the oscillations
remains constant, however, double mini-outbursts occur at later
stages of the evolution (starting at $1300$~yr). The time
separation of the two peaks in such double mini-outburst is
approximately $50$~yr. The outer part of the dead zone is not
affected by the oscillations -- it remains stationary (see
\S\ref{sect:statsol}).

\begin{figure*}
\includegraphics{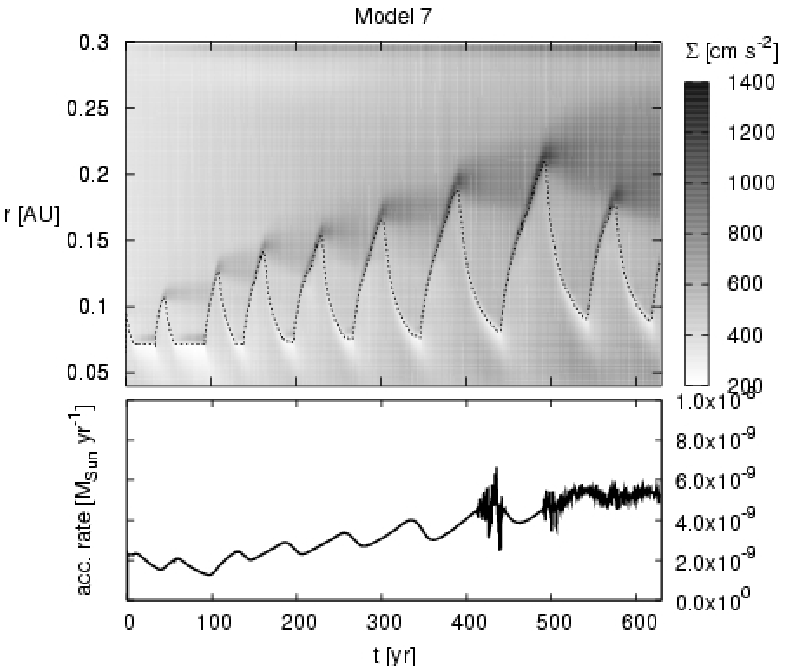}
\includegraphics{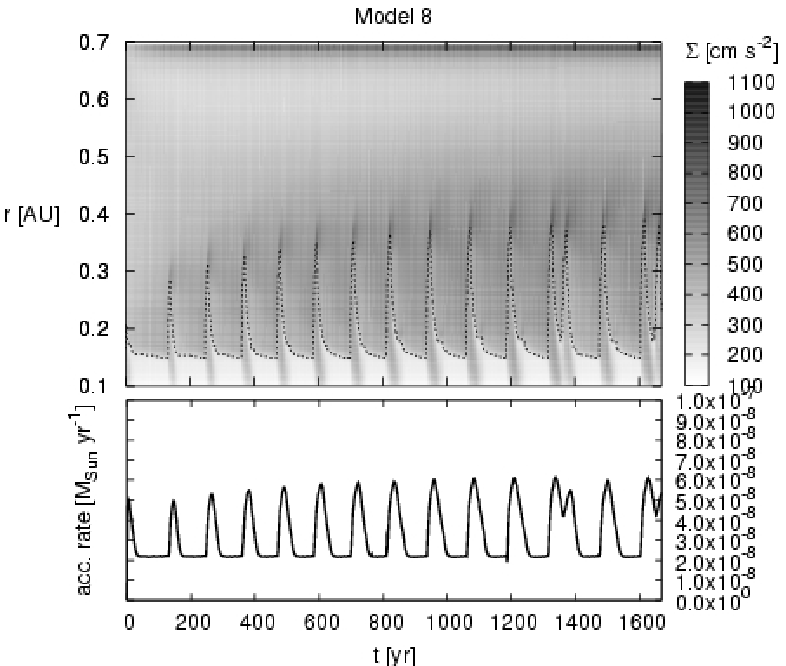}
%\vspace*{9cm}
\caption{Models 7 and 8. As in Fig. 1.}
\label{fig2}
\end{figure*}

\subsection{Summary}

In general, the numerical models exhibit two most remarkable
phenomena. If $\alpha_\mathrm{DZ} = 0$ (models 1 and 4), the dead
zone tends to split into rings. On the other hand, if
$\alpha_\mathrm{DZ} = 0.1 \alpha_\mathrm{a}$ (models 3 and 6-8), the dead
zone remains unsplit, but the inner boundary of the dead zone
starts to oscillate. The mass accumulated in this part of the dead
zone is periodically depleted. These oscillations are accompanied
by the variations in the accretion rate in the inner active disc.
Models 2 and 5 (where $\alpha_\mathrm{DZ} = 0.01 \alpha_\mathrm{a}$ are the
transitions between the two cases.

Since the ring instability observed in models 1 and 4 (and
partially also 2 and 5) was described in detail in Paper~I, in the
following we concentrate on the description and explanation of the
oscillations.

% }}}

\section{Oscillations of the inner boundary of the dead zone }\label{sec:oscil} % {{{

If $\alpha_\mathrm{DZ} \neq 0$ some accretion in the dead zone occurs.
However, as long as the accretion in the active layers dominates, the
accretion rate is, in general, a function of radius. It leads to the
accumulation of the mass and the growth of the dead zone surface
density $2\Sigma_\mathrm{DZ}$ at a rate

\begin{equation}\label{sdz_growth}
2\dot{\Sigma}_\mathrm{DZ} = \frac{1}{2\pi r}
\frac{\partial\dot{M}}{\partial r} \ ,
\end{equation}
where $\dot{M}$ is the $r$-dependent accretion rate given by the
momentum conservation law
\begin{equation}\label{am_cons}
\dot{M} = 12 \pi r^{1/2} \frac{\partial }{\partial r} \left(
\nu_\mathrm{a} \Sigma_\mathrm{a} + \nu_\mathrm{DZ} \Sigma_\mathrm{DZ}
\right) \ ,
\end{equation}
where $\Sigma_\mathrm{a}$, $\nu_\mathrm{a}$ and $\nu_\mathrm{DZ}$ are surface density of
the active layer, viscosity in the active layer and in the dead zone,
respectively. Following the standard approach of vertically
averaging we assume the viscosities in forms
\begin{equation}\label{viscsm}
\nu_{\mathrm{a,DZ}} = \alpha_{\mathrm{a,DZ}}\frac{c_\mathrm{s,m}^2}{\Omega}
\end{equation}
where $c_\mathrm{s,m} \equiv \frac{k_\mathrm{B} T_\mathrm{m}}{\mu m_\mathrm{H}}$ is the sound
speed in the mid-plane of the disc.

The vertical temperature profile of a disc with z-dependent
dissipation rate was obtained by \citet{hubeny90}. In an optically
thick approximation it can be written as

\begin{equation}\label{thubeny}
T^4(m) = \frac{3}{4} T_e^4 \left[ \tau(m)
- \tau_{\theta}(m)\right]
\end{equation}
using a mass-depth coordinate $m = \int_{z}^{\infty}\rho \mathrm{d}
z'$ changing from $0$ (at the disc surface) to $\Sigma/2$ (at
the mid-plane). $\tau(m)$ is the optical depth and
\begin{equation}
\tau_{\theta}(m) = \int_{0}^{m} \kappa(m') \theta(m')
\mathrm{d}m' \ ,
\end{equation}
is the viscosity-weighted optical depth, where
\begin{equation}
\theta(m) = \int_{0}^{m} \frac{\nu(m') \mathrm{d}m'}{\nu_\mathrm{a}\Sigma_\mathrm{a} +
\nu_\mathrm{DZ}\Sigma_\mathrm{DZ}}
\end{equation}
is the viscosity weight -- a monotonically increasing function
between $\theta(0) = 0$ and $\theta(\Sigma/2) = 1$. In our model
it assumes the form
\begin{equation}\label{thetald}
\theta(m) = \left\{
\begin{array}{lcl}
\frac{\nu_\mathrm{a} m}{\nu_\mathrm{a}\Sigma_\mathrm{a} +
\nu_\mathrm{DZ}\Sigma_\mathrm{DZ}} & \mathrm{for} &
(0,\Sigma_\mathrm{a}) \\
\frac{\nu_\mathrm{a} \Sigma_\mathrm{a} +
\nu_\mathrm{DZ}(m-\Sigma_\mathrm{a})}
{\nu_\mathrm{a}\Sigma_\mathrm{a} + \nu_\mathrm{DZ}\Sigma_\mathrm{DZ}}
& \mathrm{for} & (\Sigma_\mathrm{a},\Sigma/2)
\end{array}
\right. \ .
\end{equation}
The effective temperature, $T_e$, is given by the formula
\begin{equation}\label{teff}
T_e = \frac{9}{4\sigma}\Omega^2\left(\Sigma_\mathrm{a}\nu_\mathrm{a} +
\Sigma_\mathrm{DZ}\nu_\mathrm{DZ} \right) \ ,
\end{equation}
where $\sigma$ is the Stefan-Boltzmann constant, $\Omega$ is the
Keplerian angular velocity. Using
Eq.~(\ref{thubeny})-(\ref{thetald}) and (\ref{viscsm}) we can
express the mid-plane temperature $T_\mathrm{m} \equiv T(\Sigma/2)$ as

\begin{equation}\label{Tm}
T_\mathrm{m}^4 = \frac{3}{8} \kappa T_e^4 \frac{\alpha_\mathrm{a}
\Sigma_\mathrm{a}^2 +
\alpha_\mathrm{DZ} \Sigma_\mathrm{DZ} (\Sigma_\mathrm{DZ} + 2\Sigma_\mathrm{a})}
{\alpha_\mathrm{a}\Sigma_\mathrm{a} + \alpha_\mathrm{DZ} \Sigma_\mathrm{DZ}} \ .
\end{equation}

Further, we assume that the opacity is a piecewise power-law
function of temperature, $\kappa = \kappa_0 T_\mathrm{m}^{b}$, where the
coefficients $\kappa_0$ and $b$ for different temperature ranges
are taken from \citet{bl94}. Inserting Eq.~(\ref{Tm}) into
(\ref{am_cons}) we obtain a formula for the accretion rate of the
layered disc

\begin{eqnarray}\label{dM_ld}
\dot{M} & = & 12\pi\left( \frac{27\kappa_0}{32\sigma}
\right)^{\frac{1}{3-b}} \left( \frac{k_\mathrm{B}}{\mu
m_\mathrm{H}} \right)^{\frac{4-b}{3-b}} r^{1/2} \nonumber\\ & &
\frac{\partial }{\partial r}\left\{ \Omega^{\frac{-2+b}{3-b}}
r^{1/2} \left(\alpha_\mathrm{a} \Sigma_\mathrm{a} +
\alpha_\mathrm{DZ}\Sigma_\mathrm{DZ}\right) \right. \\ & & \left.
\left[ \alpha_\mathrm{a}\Sigma_\mathrm{a}^2 + \alpha_\mathrm{DZ}\Sigma_\mathrm{DZ}
\left(\Sigma_\mathrm{DZ} + 2\Sigma_\mathrm{a}\right)
\right]^{\frac{1}{3-b}} \right\} \ , \nonumber
\end{eqnarray}
which will be used in \S\ref{sect:statsol}.

As more and more mass accumulates in the dead zone, the mid-plane
temperature of the disc grows and when it reaches $T_\mathrm{lim}$
the disc is made active. We can estimate the surface density at
the moment of ignition, $\Sigma_\mathrm{ign}$, by substituting
$T_\mathrm{lim}$ for $T_\mathrm{m}$ in Eq.~(\ref{Tm}) and using a
metal-grains dominated opacity with $\kappa_0 = 0.1$ and $b=0.5$
(appropriate for temperatures between $203$~K and
$2290\rho^{2/49}$~K). We get

\begin{equation}
\Sigma_\mathrm{ign} = 2\left( \frac{320\sigma}{27}\frac{\mu m_\mathrm{H}}
{k_\mathrm{B}\Omega\alpha_\mathrm{DZ}}T_\mathrm{lim}^{5/2}
- \frac{\alpha_\mathrm{a} -
  \alpha_\mathrm{DZ}}{\alpha_\mathrm{DZ}}\Sigma_\mathrm{a}^2\right)^{1/2} \ .
\end{equation}

On the other hand, there is a minimum surface density
$\Sigma_\mathrm{mtn}$ necessary to maintain the disc active. It is
the surface density at which the mid-plane temperature of a fully
active disc (i.e. of a disc in which the dead zone vanishes) is
equal to $T_\mathrm{lim}$. We estimate it with the help of the
standard $\alpha$-disc solution, in which the mid-plane
temperature $T_\mathrm{m,\alpha}$ and the effective temperature
$T_{e,\alpha}$ obey the relation
\begin{equation}\label{Tmad}
T_\mathrm{m,\alpha}^4 = \frac{3}{8} \tau T_{e,\alpha}^4  \ ,
\end{equation}
where $\tau = \kappa \Sigma/2$ is the optical depth at the
mid-plane. In the same solution the effective temperature is
explicitly given by
\begin{equation}\label{Teffad}
T_{e,\alpha} = \left( \frac{9}{8\sigma} \Sigma \nu_\mathrm{a} \Omega^2 \right)^{1/4} \ .
\end{equation}
Combining Eqs.~(\ref{Tmad}) and (\ref{Teffad}) with the opacity
formula we get

\begin{eqnarray}
\Sigma_\mathrm{mtn} = \left(\frac{1280\sigma \mu m_\mathrm{H}}
{27 k_\mathrm{B} \alpha_\mathrm{a} \Omega}\right)^{1/2}
T_\mathrm{lim}^{5/4} \ .
\end{eqnarray}

\begin{figure}
\includegraphics{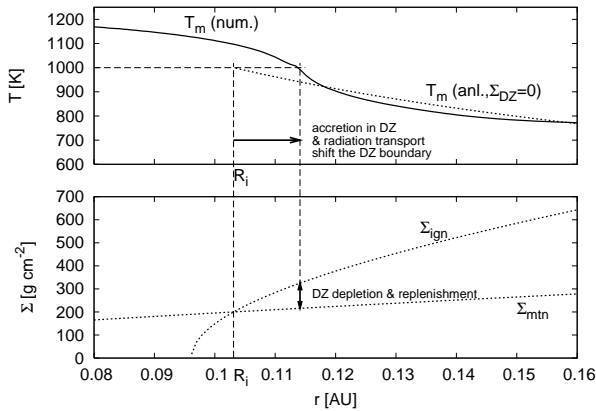}
\caption{The mechanism of the oscillations of the inner boundary of
the dead zone. Top: the mid-plane temperature of the numerical model
compared to the analytical one calculated for a disc with the empty
dead zone (i.e. $\Sigma_\mathrm{DZ}=0$). Bottom: the profiles of
$\Sigma_\mathrm{ign}$ and $\Sigma_\mathrm{mtn}$. During mini-outbursts
the surface density in the inner part of the dead zone oscillates
between these two curves.}
\label{oscmech}
\end{figure}

The inner boundary of the dead zone oscillates for reasons
explained Fig.~{\ref{oscmech}}. $\Sigma_\mathrm{ign}$ and
$\Sigma_\mathrm{mtn}$ intersect at $R_i$ defined by Gammie as the
inner boundary of the dead zone -- radius at which limit
layered disc with $\Sigma_\mathrm{DZ} = 0$ has the mid-plane
temperature $T_\mathrm{lim}$

\begin{equation}
R_i = \left( \frac{27 k_\mathrm{B} \alpha_\mathrm{a}}{320 \sigma \mu
m_\mathrm{H}}\right)^{2/3} (GM)^{1/3} \Sigma_\mathrm{a}^{4/3}
T_\mathrm{lim}^{-5/3} \ .
\end{equation}
In numerical models, however, the inner boundary of the dead zone
is shifted outwards with respect to $R_i$. This is because of two
effects that increase the temperature in the inner part of the
layered disc: the radiation transfer from the adjacent hotter
active disc (see also footnote in \citealt{gammie96}); and the
residual viscous dissipation in the dead zone.

As soon as the surface density $\Sigma$ exceeds
$\Sigma_\mathrm{ign}$, the dead zone is ignited. The ignition
always occurs at the inner boundary of the dead zone, because
$\dot{\Sigma}_\mathrm{DZ}$ is the highest there (it can be shown
by solving Eq.~\ref{sdz_growth} numerically) and
$\Sigma_\mathrm{ign}$ is an increasing function of the radius.
Therefore, the dead zone is ignited from inwards.

\begin{figure}
\includegraphics{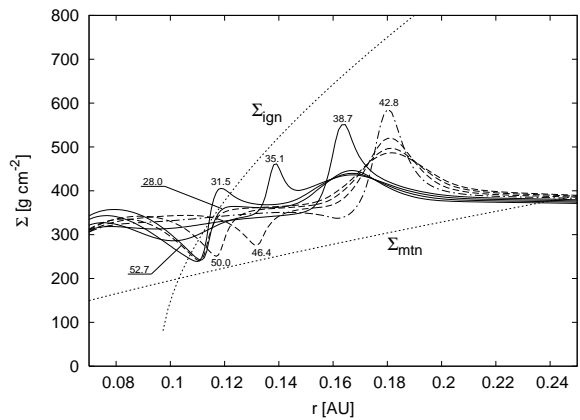}
\caption{A time-sequence of surface density profiles during the
mini-outburst in model~6 that occurs between $28$ and $53$~yr. Solid
lines ($28-38.7$~yr): the boundary of the dead zone receding from the
star; dot-dashed line ($42.8$~yr) the boundary has reached the maximum
distance from the star; dashed lines ($46.4-52.7$~yr): the boundary
approaches the star.}
\label{mob}
\end{figure}

In the ignited area the increasing viscosity causes temperature
and accretion rate to grow. The radiation flux from the ignited
region heats up and ignites the adjacent annulus of the dead zone,
so that the active area spreads outwards. On the other hand, the
increasing accretion rate tends to decrease the surface density of
the ignited region, since the mass is transported from the outer
disc at a constant rate independent on whether the dead zone has
ignited or not. When $\Sigma$ drops below $\Sigma_\mathrm{mtn}$
the ignited region becomes dead, the layered structure is
reestablished, and the accumulation of mass starts again. This
process is illustrated in Fig.~\ref{mob} which shows a time
sequence of surface-density profiles during one such
"mini-outburst" observed in the numerical model. As one might
expect, this behaviour is qualitatively similar to changes in the
surface density profile of discs in cataclysmic variables (CVs)
during an outburst; see e.g. \citet{mo85}. However, while in
CVs two types of outbursts occur (outside--in and inside--out),
only the latter seems to be possible in layered discs. As we
already mentioned, this is because at smaller radii the surface
density of the dead zone increases faster. In principle, it is
conceivable that an outburst might produce a density profile with
which the next ignition would occur away from the inner edge. We
cannot entirely exclude such possibility, although in our
simulations, covering a few hundred outbursts in total, such a
case has never been observed.

\begin{figure}
\includegraphics{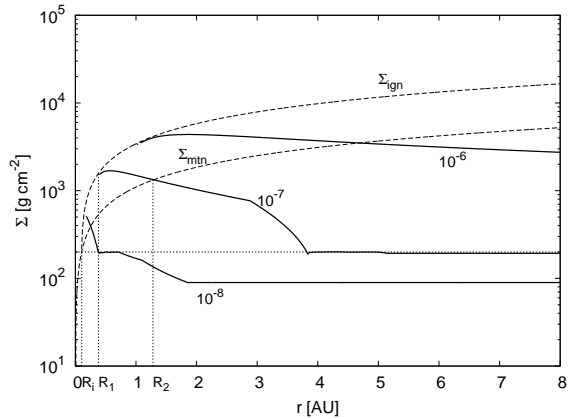}
\caption{The stationary profiles of the surface density for
different accretion rates at the inner boundary. The sharp changes
of slopes are due to switching between different opacity regimes
at temperatures $203$~K and $167$~K. The dashed lines show
$\Sigma_\mathrm{ign}$ and $\Sigma_\mathrm{mtn}$, the horizontal
dotted line denotes a value of $2\Sigma_\mathrm{a}$. The distinctive points
$R_i$, $R_1$ and $R_2$ for $\dot{M}=10^{-7}$~M$_\odot$\,yr$^{-1}$
model are denoted by the vertical dotted lines.} \label{stat}
\end{figure}

\subsection{Stationary solution}\label{sect:statsol}

The surface density of the dead zone evolves according to
Eq.~(\ref{sdz_growth}). As $\Sigma_\mathrm{DZ}$ grows and the
accretion in the dead zone becomes more and more important, the
accumulation rate $\dot{\Sigma}_\mathrm{DZ}$ decreases. At larger
radii a stationary state with radially constant accretion rate may be
reached.

We search for the stationary surface density profile
$\Sigma_\mathrm{stat}$ by
solving Eq.~(\ref{sdz_growth}) numerically with a 1-D code similar to
that described in \citet{alp01} and \citet{stepinski99}. The
computational domain ($0.1-10$~AU) is divided into 500 cells
equidistantly, and the surface density $\Sigma$ is set to some initial
value (e.g. $2\Sigma_\mathrm{a}$). Then, the the mid-plane temperature is
calculated from Eq.~(\ref{Tm}) for every grid cell and it is checked
if the cell belongs to the layered part of the disc (if $T_\mathrm{m} <
T_\mathrm{lim}$ and $\Sigma > 2\Sigma_\mathrm{a}$) or to the standard
$\alpha$-disc (otherwise). In the the later case, the temperature is
recomputed according to Eq.~(\ref{Tmad}) to obtain the consistent
$\alpha$-disc solution there.

The opacity coefficients $\kappa_0$ and $b$ are determined using
the first three Bell \& Lin regimes only since the analytical
solution does not always exist for the remaining ones (i.e. we
take into account ice dominated opacity with $\kappa_0=2\times
10^{-4}$ and $b=2$ below $167$~K, the jump due to ice evaporation
with $\kappa_0=2\times 10^{16}$ and $b=-7$ between $167$ and
$203$~K, and metal grains dominated opacity with $\kappa_0=0.1$
and $b=0.5$ between $203$ and $2290\rho^{2/49}$~K). Our opacity
formula may be incorrect for the part of the inner $\alpha$-disc
(and, occasionally, also for the several cells of the layered
region where the temperature exceeds $2290\rho^{2/49}$~K).
However, using this toy model we are not particularly interested
in the very inner disc, which is much better described by the
TRAMP simulation anyway. Moreover, recent works on opacity in
protoplanetary discs (see e.g. \citealt{shhis03}) show that the
metal grains evaporate at a slightly higher temperature than that
adopted by \citet{bl94}.

Subsequently, the accretion rate $\dot{M}$ is computed at the
boundaries of each cell from Eq.~(\ref{dM_ld}) for the layered disc or
from
\begin{equation}
\dot{M}_{\alpha} = 3\pi\nu_\mathrm{a} \Sigma
\end{equation}
for the $\alpha$-disc. Finally, the amount of mass proportional to
some small time-step $dt$ is transported between cells and
$\Sigma$ is updated. The accretion rate $\dot{M}_\mathrm{OB}$ at
the outer boundary of the domain is a free parameter, and a
free outflow condition is imposed at the inner boundary. The
computation ends when the accretion rate throughout the
computational domain converges to $\dot{M}_\mathrm{OB}$, i.e. when
the stationary state is reached. Typically, it happens after
several times $10^{5}$~yr depending on $\dot{M}_\mathrm{OB}$.

The stationary layered disc surface density profiles
$\Sigma_\mathrm{stat}$ for $\dot{M} = 10^{-8}$, $10^{-7}$ and
$10^{-6}$ are shown in Fig.~{\ref{stat}}. They end at the radius
$R_1$ where they cross $\Sigma_\mathrm{ign}$ curve, since the
stationary layered disc solution does not exist below $R_1$. The
radius $R_2$ at which $\Sigma_\mathrm{stat}$ curve crosses
$\Sigma_\mathrm{mtn}$ is another distinctive point: it separates
the region of the dead zone which after the ignition is able to
maintain the temperature above $T_\mathrm{lim}$ from the one which
is not. The dead zone ends at radius $R_i$ where the surface
density drops below $2\Sigma_\mathrm{a}$.

In this manner, the dead zone can be divided into three parts: a) the
non-stationary periodically depleted and replenished part between
$R_i$ and $R_1$, b) the stationary, but "combustible" part between
$R_1$ and $R_2$, and c) the stationary "incombustible" part between
$R_2$ and $R_o$.

An upper limit for the mass accreted during one mini-outburst in
region a) can be obtained by integration the difference between
$\Sigma_\mathrm{ign}$ and $\Sigma_\mathrm{mtn}$ curves there

\begin{equation}
M_\mathrm{a} = 2\pi \int_{R_i}^{R_1}
r(\Sigma_\mathrm{ign} - \Sigma_\mathrm{mtn}) dr \ .
\end{equation}
Taking values $R_1$ from the numerical model $R_1 = 0.15, 0.38$ and
$0.96$~AU for $\dot{M} = 10^{-8}, 10^{-7}$ and
$10^{-6}$~M$_\odot$\,yr$^{-1}$, respectively, we get

\begin{equation}
M_\mathrm{a} \approx \left\{
\begin{array}{lll}
7.2\times 10^{-7}\mathrm{M}_\odot           & \mathrm{for}\ \dot{M} = 10^{-8}\ \mathrm{M}_\odot\,\mathrm{yr}^{-1} \\
2.6\times 10^{-5}\mathrm{M}_\odot & \mathrm{for}\ \dot{M} = 10^{-7}\ \mathrm{M}_\odot\,\mathrm{yr}^{-1} \\
5.2\times 10^{-4}\mathrm{M}_\odot              & \mathrm{for}\ \dot{M} = 10^{-6}\ \mathrm{M}_\odot\,\mathrm{yr}^{-1}
\end{array}
\right.
\end{equation}

The amount of mass available for the accretion in the region b) may
be interesting in relation to FU~Orionis type outbursts. It can be
determined as
\begin{equation}
M_b = 2\pi \int_{R_1}^{R_2}
r(\Sigma_\mathrm{stat} - \Sigma_\mathrm{mtn}) dr \ .
\end{equation}
Integrating numerically the stationary solutions obtained by 1-D model
we get
\begin{equation}
M_b \approx \left\{
\begin{array}{lll}
2.0\times 10^{-6} \mathrm{M}_\odot           & \mathrm{for}\ \dot{M} = 10^{-8}\ \mathrm{M}_\odot\,\mathrm{yr}^{-1} \\
2.7\times 10^{-4} \mathrm{M}_\odot & \mathrm{for}\ \dot{M} = 10^{-7}\ \mathrm{M}_\odot\,\mathrm{yr}^{-1} \\
0.01 \mathrm{M}_\odot              & \mathrm{for}\ \dot{M} = 10^{-6}\ \mathrm{M}_\odot\,\mathrm{yr}^{-1}
\end{array}
\right.
\end{equation}

The typical amount of mass accreted during the FU~Ori type outburst
is $\sim 0.01$~M$_\odot$ \citep{hk96} which is comparable to the
mass stored and available for the accretion in the dead zone in the
case with $\dot{M} = 10^{-6}$~M$\odot$\,yr$^{-1}$.

% }}}

\section{Discussion} % {{{

The two-dimensional radiation-hydrodynamic simulations of the
layered disc we carried out showed that the behaviour of the disc
strongly depends on the presence of a small residual viscosity in
the dead zone $\alpha_\mathrm{DZ}$. In the models with
$\alpha_\mathrm{DZ} = 0$ the ring instability develops as
described in Paper~I. The  rings may eventually decay which would
temporarily increase the accretion rate onto the central star by a
factor of several. On the other hand, in the models with
$\alpha_\mathrm{DZ}$ set to 10\% of the viscosity in the active
parts of the disc, the dead zone remains contiguous, but its inner
boundary oscillates.

These oscillations are more or less regular with period between
several tens and several hundreds years depending on model
parameters. They can be accompanied by variations of the accretion
rate on a time-scale as short as ten years, at a relative amplitude
reaching from a few per cent to 200 -- 300 per cent in the high
viscosity Model 8. These variations would be rather difficult to
observe, however it does not seem absolutely impossible. Since the
properties of these oscillations are quite sensitive to physical
conditions in the disc, such observations could help to constrain
the set of models of protoplanetary discs.

Similar variations of the accretion rate were also found by
\citet{stepinski99} using an one-dimensional hydrodynamic code.
However, the period was much longer than in our model
($10^3-10^4$~yr). The reason for that may be that in his models
the ignition of the inner part of the dead zone is much more
difficult due to the absence of the radiation transport (and,
generally speaking, due to oversimplified description of the
interactions between the inner hot $\alpha$-disc and the layered
region).

We give an analytical description of these oscillations based on the vertically
averaged model modified for the case of layered disc. It illustrates the
mechanism of the dead zone oscillations and tests the numerical model.

Finally, we perform a simple 1-D model of the layered disc based on the
analytical solution to follow the evolution of the whole layered region for a
long time. We found that with $\alpha_\mathrm{DZ} \neq 0$ a stationary layered
region may exist at higher radii (this was also observed in numerical Model~8).
Part of this stationary region is still combustible, i.e. it may be made active
by increasing the temperature by some external event, and part of its mass is
available for accretion.

We compare the amount of mass stored in the combustible region to the mass
accreted during a FU~Ori type outburst following the idea that the dead zone
can serve as a mass reservoir for such event. We found, that these masses are
comparable for the case of massive disc with the high accretion rate, which is
in agreement with the observation that the FU~Ori objects are very young
systems.

% }}}

\section{Acknowledgments}

This research was supported in part by the European Community's
Human Potential Programme under contract HPRN-CT-2002-00308,
PLANETS. RW acknowledges financial support by the Grant Agency of
the Academy of Sciences of the Czech Republic under the grants
AVOZ 10030501 and B3003106. AG and MR were supported by the Polish
Ministry of Science through the grant No. 1 P03D 026 26.

 % }}}

\label{lastpage}


\begin{thebibliography}{} % {{{

\bibitem[\protect\citeauthoryear{Armitage, Livio \& Pringle}{Armitage et~al.}{2001}]{alp01}
Armitage P. J., Livio M. and Pringle J. E., 2001, MNRAS, 324, 705

\bibitem[\protect\citeauthoryear{Balbus \& Hawley}{1991a}]{bh91a}
Balbus S. A., Hawley J. F., 1991a, ApJ, 376, 214

\bibitem[\protect\citeauthoryear{Balbus \& Hawley}{1991b}]{bh91b}
Balbus S. A., Hawley J. F., 1991b, ApJ, 376, 223

\bibitem[\protect\citeauthoryear{Bell \& Lin}{1994}]{bl94}
Bell K. R., Lin D. N. C., 1994, ApJ, 477, 987

\bibitem[\protect\citeauthoryear{Cannizzo}{2001}]{cannizzo01}
Cannizzo, J. K., 2001, ApJ, 556, 847

\bibitem[\protect\citeauthoryear{Fleming \& Stone}{2003}]{fs03}
Fleming T., Stone J. M., 2003, ApJ, 585, 908

\bibitem[\protect\citeauthoryear{Fromang, Terquem \& Balbus}{Fromang et~al.}{2002}]{ftb02}
Fromang S., Terquem C., Balbus S. A., 2002, MNRAS, 329, 18

\bibitem[\protect\citeauthoryear{Gammie}{1996}]{gammie96}
Gammie C. F., 1996, ApJ, 457, 355

\bibitem[\protect\citeauthoryear{Glassgold, Najita \& Igea}{Glassgold et~al.}{1997}]{gni97}
Glassgold A. E., Najita J., Igea J., 1997, ApJ, 480, 344

\bibitem[\protect\citeauthoryear{Hartmann \& Kenyon}{1996}]{hk96}
Hartmann L. Kenyon S. J., 1996, Annu. Rev. Astron. Astrophys., 34, 207

\bibitem[\protect\citeauthoryear{Hawley, Gammie \& Balbus}{1995}]{hawley95}
Hawley, J. F., Gammie, C. F., Balbus S. A., 1995, ApJ, 440, 742

\bibitem[\protect\citeauthoryear{Hubeny}{1990}]{hubeny90}
Hubeny I., 1990, ApJ, 351, 632

\bibitem[\protect\citeauthoryear{Klahr, Henning \& Kley}{Klahr et~al.}{1999}]{khk99}
Klahr H. H., Henning T., Kley W., 1999, ApJ, 514, 325

\bibitem[\protect\citeauthoryear{Kley \& Lin}{1999}]{kl99}
Kley W., Lin D. N. C., 1999, ApJ, 518, 833

\bibitem[\protect\citeauthoryear{Lodato \& Clarke}{2004}]{lc04}
Lodato G., Clarke C. J., 2004, MNRAS, 353, 841

\bibitem[\protect\citeauthoryear{Meyer \& Meyer-Hofmeister}{1981}]{mm-h81}
Meyer F., Meyer-Hofmeister E., 1981, A\&A, 104, L10

\bibitem[\protect\citeauthoryear{Mineshige \& Osaki}{1985}]{mo85}
Mineshige, S., Osaki, Y., 1985, PASJ, 37, 1

\bibitem[\protect\citeauthoryear{Papaloizou \& Pringle}{1985}]{pp85}
Papaloizou J. C. B., Pringle J. E., 1985, MNRAS, 213, 799

\bibitem[\protect\citeauthoryear{Semenov et~al.}{2003}]{shhis03}
Semenov D., Henning Th., Helling Ch., Ilgner M., Sedlmayr E., 2003, A\&A, 410, 611

\bibitem[\protect\citeauthoryear{Shakura \& Sunayev}{1973}]{ss73}
Shakura N. I., Sunyaev R. A., 1973, A\&A, 24, 337

\bibitem[\protect\citeauthoryear{Stepinski}{1999}]{stepinski99}
Stepinski T., 1999, 30th Annual Lunar and Planetary Science Conference, Houston, TX, abstract no. 1205

\bibitem[\protect\citeauthoryear{Stone et al.}{1996}]{stone96}
Stone, J. M., Hawley, J. F., Gammie, C. F., Balbus S. A., 1996,
ApJ, 463, 656

\bibitem[\protect\citeauthoryear{Umebayashi}{1983}]{umebayashi83}
Umebayashi T., 1983, Prog. Theor. Phys., 69, 480

\bibitem[\protect\citeauthoryear{Umebayashi \& Nakano}{1981}]{un81}
Umebayashi T., Nakano T., 1981, PASJ, 33, 617

\bibitem[\protect\citeauthoryear{W\"unsch, Klahr \& R\'o\.zyczka}{W\"unsch et~al.}{2005}]{wkr05}
W\"unsch R., Klahr, H. H., R\'o\.zyczka M., 2005, MNRAS, 362, 361

\end{thebibliography}
\end{document}